\PassOptionsToPackage{dvipsnames,table,xcdraw}{xcolor}
\documentclass[preprints,article,accept,pdftex,moreauthors]{Definitions/mdpi} 
\firstpage{1} 
\pubvolume{1}
\issuenum{1}
\articlenumber{0}
\pubyear{2025}
\copyrightyear{2024}
\datereceived{ } 
\daterevised{ } 
\dateaccepted{ } 
\datepublished{ } 
\hreflink{https://doi.org/} 

\usepackage{CJKutf8}
\usepackage{tabularx}

\usepackage{caption}
\Title{Using LLMs to Infer Non-Binary COVID-19 Sentiments of Chinese Micro-bloggers}

\TitleCitation{Using LLMs to Infer Non-Binary COVID-19 Sentiments of Chinese Micro-bloggers}

\Author{Jerry Chongyi Hu $^{1,\dagger}$, Mohammed Shahid Modi $^{1,\dagger}$ and Boleslaw K. Szymanski $^{1,\dagger}$*}

\AuthorNames{Jerry Chongyi Hu, Mohammed Shahid Modi and Boleslaw K. Szymanski}

\AuthorCitation{Hu, J.C., Modi, M.S., \& Szymanski, B.K.}

\address{%
$^{1}$ \quad Department of Computer Science and Network Science and Technology Center, Rensselaer Polytechnic Institute, Troy, New York, USA}

\corres{Correspondence: szymab@rpi.edu}

\firstnote{These authors contributed equally to this work.}  

\abstract{Studying public sentiment during crises is crucial for understanding how opinions and sentiments shift, resulting in polarized societies. We study Weibo, the most popular microblogging site in China, using posts made during the outbreak of the COVID-19 crisis. The study period includes the pre-COVID-19 stage, the outbreak stage, and the early stage of epidemic prevention. We use Llama 3 8B, a Large Language Model, to analyze users' sentiments on the platform by classifying them into positive, negative, sarcastic, and neutral categories. Analyzing sentiment shifts on Weibo provides insights into how social events and government actions influence public opinion. This study contributes to understanding the dynamics of social sentiments during health crises, fulfilling a gap in sentiment analysis for Chinese platforms. By examining these dynamics, we aim to offer valuable perspectives on digital communication's role in shaping society's responses during unprecedented global challenges.}

\begin{document}

\section{Introduction}

The COVID-19 pandemic was one of the most disruptive events in recent history, and its effect on social media was profound. It caused widespread panic and led to the proliferation of misinformation and fake news through media outlets, the internet, and social networks \cite{caceres2022impact}. It caused measurable differences in the sentiments and emotions expressed by social media users through their posts, including on the Chinese microblogging website Weibo \cite{wu2022classification}. 

In this paper, we analyze user sentiments on the Weibo platform during the initial five-month period during which the pandemic started to spread in China (November 2019 - March 2020). We categorize millions of Chinese posts into four sentiment categories (positive, negative, neutral, and sarcastic) using the Meta Llama 3 8B Instruction-tuned variant LLM \cite{llama3modelcard}. Then, we analyze them to describe the sentiment dynamics on Weibo.

While popular social media such as Facebook reported removing large numbers of posts and users to prevent the spread of misinformation and fake accounts during the pandemic \cite{fb2021mods}, it is unclear to what extent Chinese social media Weibo moderated posts during this period. Weibo's content moderation system is community-driven and has several unique features. For instance, some Weibo users can voluntarily report posts that promote unsocial behavior. The platform treats such reports differently, and a panel of community jurors is sometimes selected from the platform's users to make case decisions instead of relying just on platform administration \cite{hu2023rule}.

Thus, it is essential to consider how moderation and censorship may have affected users' present sentiments, as users may have self-censored or masked sensitive speech to avoid community reporters or AI-based detection. One way they may have done this is by using sarcasm; we measure sarcastic posts independent of positive or negative sentiments. This work's novel contributions are (1) a technique for large-scale Chinese text sentiment classification using an LLM instead of standard NLP techniques and (2) sentiment dynamics analysis of pandemic-time Weibo posts.

The following section reviews some literature about Weibo, sentiment analysis, and LLM-based classification. Next, we describe our Methods, including our dataset and few-shot prompting approach. We then present our results and discuss possibilities for future work.

\subsection{Related Work}

Studies, like \cite{caceres2022impact}, have been conducted about the impact of misinformation during the COVID-19 pandemic. These studies show that social media, mainstream media, and other sources of information inadvertently propagate misinformation or false information. This misinformation likely contributed to increased negative or sarcastic tones in social media posts.

A study on perceptions of government legitimacy and how they affected individual willingness to disclose private information to the government in Shanghai, China \cite{meng2024pathos} found that moral legitimacy took precedence over cognitive legitimacy, with altruism being the mediator. The study states that Chinese moral and collective values played an important role in privacy disclosures, and this finding may help explain sentiment dynamics on Weibo. In addition, \cite{zhang2022avoid} showed that Weibo users used memes to avoid online censorship. Such use of memes supports our decision to track sarcasm as a sentiment/tone, as Weibo users may have attempted to avoid moderation. 

In \cite{luo2021vac}, the authors examined 700,000 Twitter posts and 400,000 Weibo posts over three months from December 2020 to February 2021. The study compared sentiments about COVID-19 vaccines on Twitter and Weibo and found that neutral posts were most common on Twitter. At the same time, positive tweets were most common on Weibo, among other key observations. In contrast, our study examines the broader sentiment on Weibo during the initial months of the pandemic by analyzing millions of Weibo posts.

Lexicon-based approaches are commonly used for sentiment classification due to their effectiveness over supervised approaches \cite{lukas2014simpler}. Large Language models have also emerged as effective tools for context-rich sentiment analysis \cite{ghatora2024sentiment}. We use an LLM for sentiment classification in our study.

\section{Materials and Methods}

\begin{figure}[H]

\centering

\includegraphics[width=0.55\textwidth]{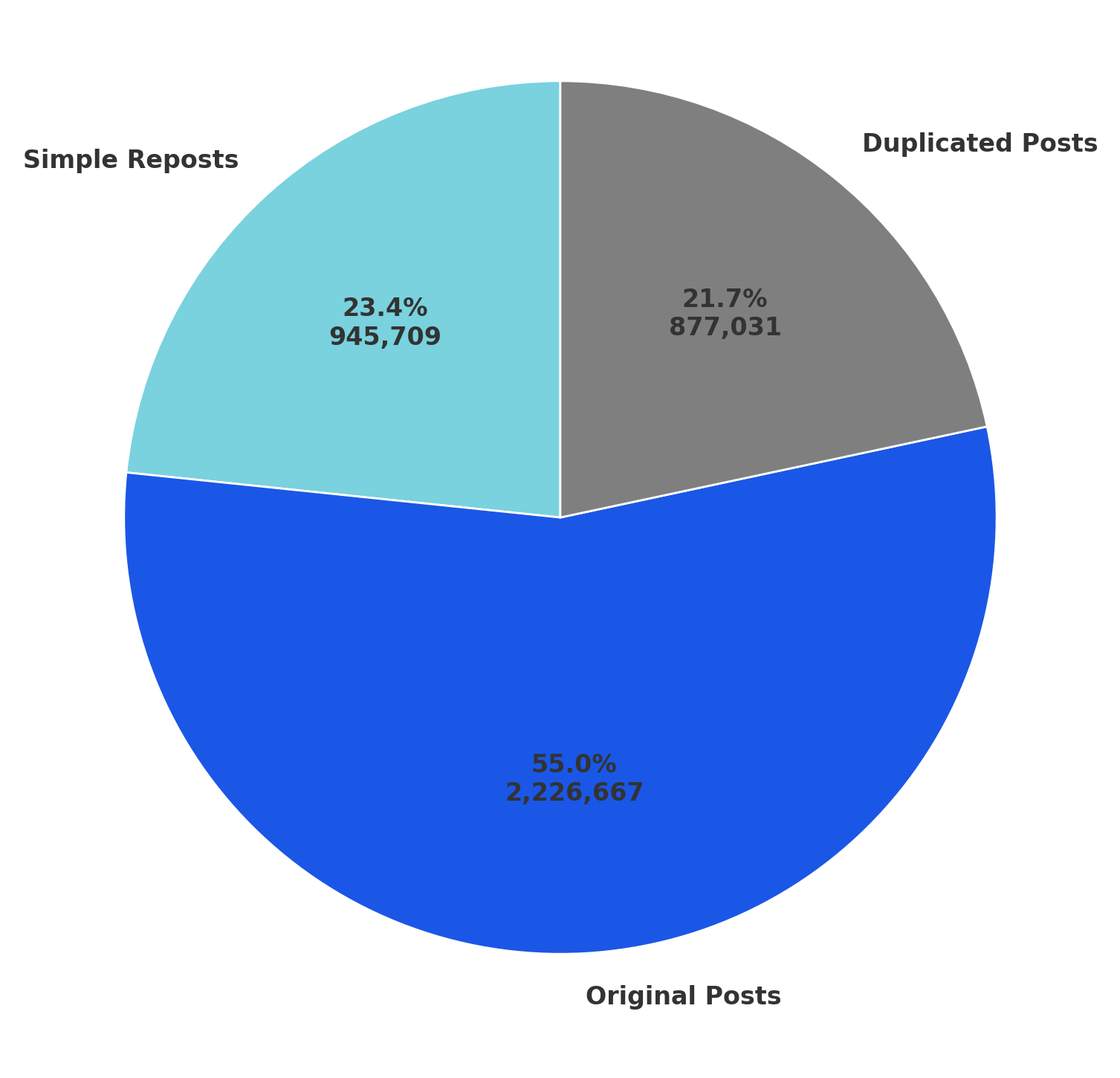}

\caption{Distribution of duplicate, original posts and reposts. 21.7\% of the posts are duplicated (877,031 posts), 55\% are distinct posts (2,226,667 posts), and 23.4\% of the posts are reposts (945,709 posts).}

\label{dataset_info}

\end{figure}

We use the Weibo COVID-19 Dataset from Harvard \cite{DVN/DULFFJ_2020}. The dataset contains a total of 4,049,407 posts. Based on post content, the dataset includes 2,226,667 distinct posts and 877,031 duplicated posts, duplicated content that may be human-made propaganda or bot-generated spam, and 945,709 reposts. We cleaned and categorized the dataset and presented the types and counts of posts in Figure~\ref{dataset_info}. Some of the duplicates are straightforward to identify as they are entirely identical. However, specific spam posts employ subtle variations, such as adding spaces or tabs, to avoid spam filters on Weibo. To address this issue, we developed a program to preprocess and clean the Weibo dataset, ensuring its quality and consistency before classifying.

\begin{table}[H]
\captionsetup{justification=centering}
\caption{Metadata of each post in the dataset}
\centering
\begin{tabular}{|l|l|}
\hline
\textbf{Field name} & \textbf{Field Description} \\ \hline
\rowcolor{gray!30} 
From & Device used to post \\ \hline
Content & Text content of the post \\ \hline
\rowcolor{gray!30}
Repost - Content & Text content of the repost \\ \hline
Repost - Imgs & Images in the repost \\ \hline
\rowcolor{gray!30}
Repost - Timestamp & Timestamp of the repost \\ \hline
Repost - Username & Numeric ID identifying the reposter \\ \hline
\rowcolor{gray!30} 
Timestamp & Timestamp of the post \\ \hline
User\_ID & Numeric ID identifying the poster \\ \hline
\rowcolor{gray!30} 
Weibo\_ID & Unique ID of the post \\ \hline
\end{tabular}
\label{tab:json_description}
\end{table}

We structured the dataset in JSON format. Each post has content and metadata. Table~\ref{tab:json_description} shows this metadata together with their fields.

\subsection{Post types and Time Periods}

As mentioned earlier, the dataset contains a significant number of duplicate posts. Differentiating between distinct posts, duplicate posts, and reposts is essential to improving classification accuracy. 

\begin{itemize}

\item Distinct posts: These are original posts created by users with unique content that does not replicate existing posts in the dataset. 

\item Duplicate posts: These are identical posts that appear multiple times due to content being manually created numerous times without any modifications or automated replication (e.g., spam).

\item Reposts: These occur when users share existing posts with or without their comments. Some reposts appear as distinct posts because users have added unique comments or annotations when sharing the content. Conversely, some reposts closely resemble duplicate posts, as they include minimal changes, such as default repost tags or no additional input from the user. Still, we will identify those posts as reposts and exclude them from duplicate posts, as those posts represent individuals' sentiments just as distinct posts do. We identify reposts by parsing the content of the posts and checking for the presence of "//." If the post content ends with "//," it indicates that the post is a repost of another post.

\end{itemize}

\begin{figure}[H]

\centering

\includegraphics[width=0.77\textwidth]{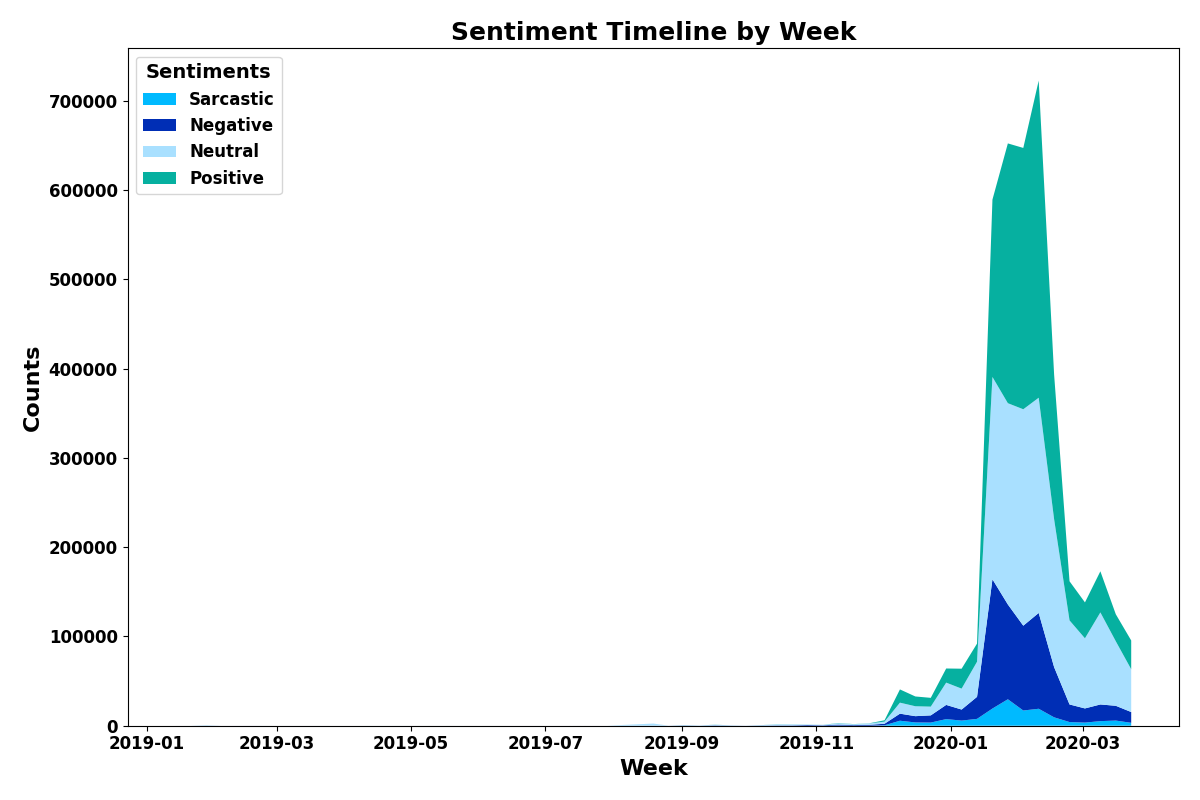}

\caption{A sentiment timeline stacked area chart illustrating weekly counts of positive, neutral, negative, and sarcastic sentiments from January 2019 to March 2020. Posts in all sentiment categories are significantly stacked around late 2019, peaking in early 2020, with positive and neutral sentiments dominating the counts.}

\label{total_stack}

\end{figure}

Most posts in the dataset were collected between December 2019 and March 2020, as shown in Figure~\ref{total_stack}, which also visualizes the posts of each sentiment. However, a small number of posts about the African Swine Fever outbreak are also present in the dataset \textbf{between January 2019 and December 2019}. 

\subsection{Few-shot Prompting}

We run the Llama 3 model for sentiment classification of Weibo posts. We use the few-shot prompting technique.
We manually identified a few posts containing each of the four sentiment types. Then, we created a prompt for the Llama 3 model to predict the sentiment of posts using five manually labeled posts as an example. The prompt we used is quoted below. For the convenience of readers not familiar with Chinese, we have translated each Weibo post to English using the Google Translate website and placed it under the Chinese version for the readers. These translations were not a part of prompting the model.

\begin{quote}

\textit{``You are a bot designed to judge Chinese sentences as having positive, negative, or neutral sentiment. I'm going to provide posts from Chinese social media. **IMPORTANT: You must only answer with "positive,” "negative,” "sarcastic," or “neutral.” Do not explain your response or include other text.**}

\textit{Use the following criteria for your judgment:}

\textit{1. If the post speaks well of a figure or event that is commonly regarded as a good figure or event, with no unnecessary exaggeration, it is 'positive' sentiment.}

\textit{2. If the post disparages a figure or event commonly regarded as a wrong figure or event, then it is a 'negative' sentiment.}

\textit{3. If the post does not use emotional language and consists of matter-of-fact reporting of factual statements, it corresponds to a 'neutral' sentiment.}

\textit{4. Sarcasm in the post, which appears with exaggerated emotion, pretend naivete, or other common sarcastic tone indicators, should correspond to a 'sarcastic' sentiment.}

\textit{IMPORTANT: Some posts may be appended with the original post that the user replied to. They use the format: [reply]//[original post]. You must predict the sentiment of the reply only, but you may use the original post as context to understand the reply.}

\textit{Here are some examples:}

\textit{Post:} \begin{CJK}{UTF8}{gbsn}\#关注新型肺炎\#【国家监委派出调查组，全面调查涉及李文亮医生有关问题】经中央批准，国家监察 委员会决定派出调查组赴湖北省武汉市，就群众反映的涉及李文亮医生的有关问题作全面调查。O国家监委派出调查组，全面调查涉及李文亮医生有关问题

\end{CJK}

\textit{Translation: \#focus on new pneumonia\# [The National Supervisory Commission dispatched an investigation team to investigate issues related to Dr. Li Wenliang comprehensively]. With the approval of the Central Committee, the National Supervisory Commission decided to send an investigation team to Wuhan City, Hubei Province, to investigate the complaints reported by the masses involving Dr. Li Wenliang's findings. The National Supervisory Commission dispatched a team of epidemiologists to evaluate these findings.} 

\noindent\textit{Expected Answer: Neutral}

\textit{Post:} \begin{CJK}{UTF8}{gbsn}妈妈 我已经快二十天没喝奶茶 没有大吃大喝了 求求你赶紧疫情结束 我人都快没了 我想上课我想喝奶茶我想吃烧烤

\end{CJK}

\textit{Translation: Mom, I haven't had milk tea or a big meal for almost 20 days. Please end the epidemic soon. I'm practically dead. I want to go to class, drink milk tea, and have a barbecue.}

\noindent\textit{Expected Answer: Negative}

\textit{Post:} \begin{CJK}{UTF8}{gbsn}高中时就暗恋他已久，高三毕业那天我鼓起勇气表白，万万没想到他居然也默默喜欢着我，填志愿时 也选择了同一个城市。后来工作了异地了四年，晃眼我们走过了十年呢，19年时我们步入了婚姻礼堂。我想 这世上最幸福的事情之一 那就是两个人都互相深爱并且坚持吧~疫情过后 春暖花开，我们想去武大看樱花。

\end{CJK}

\textit{Translation: I had a crush on him for a long time in high school. On graduation day, I mustered up the courage to confess my love. I never expected that he would secretly like me. When filling out the application form, he chose the same city. Later, I worked in a different place for four years. In the blink of an eye, we have gone through ten years. In 2019, we entered the marriage hall. One of the happiest things in the world is that two people sincerely love each other and persevere. After the epidemic, spring will come, and we want to go to Wuhan University to see the cherry blossoms.}

\noindent\textit{Expected Answer: Positive}

\textit{Post:} \begin{CJK}{UTF8}{gbsn}就你们敢说实话，//@7362410961:世卫组织说目前只有瑞德西韦可能有效，中科院双黄连有效，南京 大学说金银花有效，北京大学沐舒坦有效，南开大学说姜、大枣、龙眼肉都能预防。中国人民真幸福，这么 多常见药物、食品可以抗新型冠状病毒，还慌什么呢？\end{CJK}

\textit{Translation: You are the only ones who dare to tell the truth, //@7362410961: The World Health Organization said that only Remdesivir may be effective at present, the Chinese Academy of Sciences [stated that] Shuanghuanglian is effective, Nanjing University said that honeysuckle is effective, Peking University Mucosolvan is effective, and Nankai University said that ginger, jujube, and longan meat can prevent it. The Chinese people are so lucky. So many common medicines and foods can resist the new coronavirus. What are they still panicking about?}

\noindent\textit{Expected Answer: Sarcastic}

\textit{Post:} \begin{CJK}{UTF8}{gbsn}无言//因为疫情我猛然发现不管是新闻里还是现实生活中，从大官到小官，究竟还有多少智力缺陷人 士，干出来的事每天都让老百姓瞠目结舌 \end{CJK}

\textit{Translation: Speechless//Because of the epidemic, I suddenly realized that whether it is in the news or real life, from high officials to minor officials, how many mentally disabled people are there and what they do every day makes people dumbfounded.}

\noindent\textit{Expected Answer: Negative}

\textit{Now we begin. Classify:"}

\end{quote}

\subsection{LLM Accuracy Comparison}

We used the instruction-tuned Llama 3 model with 8 billion parameters. We ran the model locally on one of Rensselaer Polytechnic Institute's FOCI cluster nodes, specifically on an Nvidia Ampere A100 GPU. We downloaded the model from an online platform, HuggingFace, using the HuggingFace Python API.

\begin{table}[H]

\captionsetup{justification=centering}

\caption{Llama 3 Confusion Matrix}

\setlength{\tabcolsep}{0.3em} 

\renewcommand{\arraystretch}{1.5}

\centering

\begin{tabular}{|c|c|c|c|c|}

\hline

& \textbf{Sarcastic} & \textbf{Neutral} & \textbf{Negative} & \textbf{Positive} \\ \hline

\textbf{Sarcastic} & \cellcolor{green!25}3 & 3 & 1 & 0 \\ \hline

\textbf{Neutral} & 1 & \cellcolor{green!25}69 & 1 & 5 \\ \hline

\textbf{Negative} & 1 & 8 & \cellcolor{green!25}25 & 1 \\ \hline

\textbf{Positive} & 2 & 17 & 3 & \cellcolor{green!25}59 \\ \hline

\end{tabular}

\label{tab:split_diagonal}

\end{table}

We selected a subset of 199 posts from the dataset to validate the model's accuracy. We categorized them into positive, negative, neutral, or sarcastic using a more robust model with 100B parameters, OpenAI o1-mini, using the OpenAI API. A native Chinese speaker evaluated the model's predictions and found them accurate. Next, we categorized the same subset of posts using our chosen model, Llama 3 8B, and the slightly newer Llama 3.1 8B. Table~\ref{tab:split_diagonal} compares this model's confusion matrix to the o1-mini model's ground truth classifications.

\begin{table}[H]

\captionsetup{justification=centering}

\caption{Comparison of Precision, Recall, and F1 Scores of Llama 3 and 3.1}


\setlength{\tabcolsep}{0.5em} 

\centering

\begin{tabular}{|c|c|c|c|c|c|c|}

\hline 

& \multicolumn{3}{c|}{\textbf{Llama 3}} & \multicolumn{3}{c|}{\textbf{Llama 3.1}} \\

\hline

\textbf{Metric>} & \textbf{Precision} & \textbf{Recall} & \textbf{F1-score} & \textbf{Precision} & \textbf{Recall} & \textbf{F1-score} \\

\hline

Sarcastic & 0.4286 & 0.4286 & 0.4286 & 0.5714 & 0.4444 & 0.5000 \\

Neutral & 0.7113 & 0.9079 & 0.7977 & 0.7320 & 0.8353 & 0.7802 \\

Negative & 0.8333 & 0.7143 & 0.7692 & 0.9000 & 0.7105 & 0.7941 \\

Positive & 0.9077 & 0.7284 & 0.8082 & 0.7846 & 0.7612 & 0.7727 \\

\hline

\end{tabular}


\label{tab:scores}

\end{table}

Llama 3 8B achieves a weighted F1-score of 78.39\% overall on the validation set. Llama 3.1 achieves a slightly lower score of 76.88\%. Table~\ref{tab:scores} compares the per-type precision, recall, and F1 score. The table shows that Llamas 3 achieved F1 scores of around 80\% for the three well-represented sentiments in the dataset. The exception is sarcastic sentiment, which is the least frequent in the data set, and its F1 score is much lower at 42.86\%. An infrequency of appearance and complexity of expression are the likely reasons for the low F1 score. Llama 3.1 8B is more accurate than Llama 3 for the sarcastic and negative sentiment posts in terms of F1 score, but it is less accurate for the other two types. Therefore, we decided that Llama 3 8B would be the better candidate for our experiment.

\section{Results}

\begin{figure}[H]

\captionsetup[subfloat]{justification=centering}
\centering
    \subfloat[]{
        \includegraphics[width=0.4\textwidth]{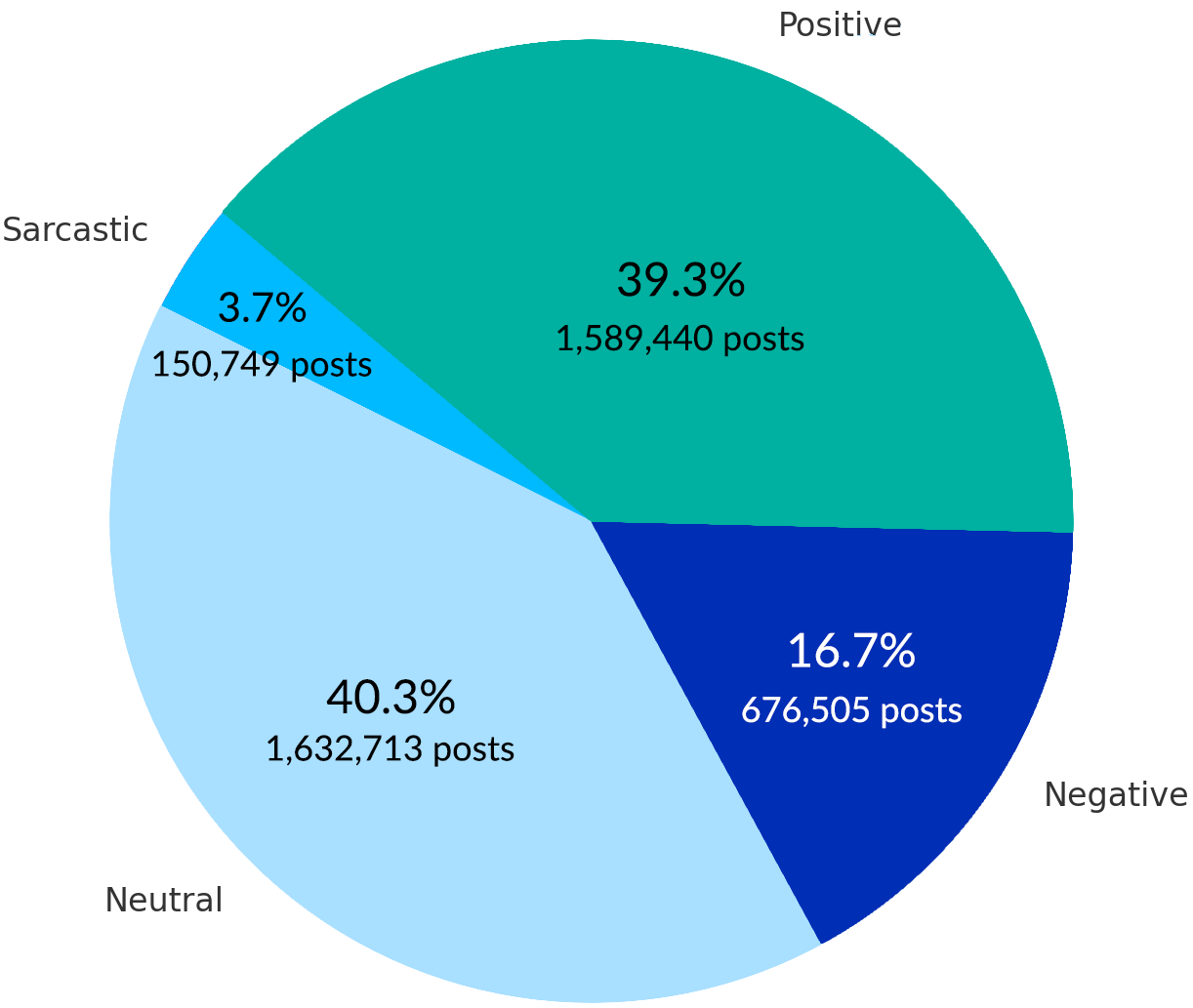}
        \label{piechart:a}}
    \hfil
    \subfloat[]{
        \includegraphics[width=0.4\textwidth]{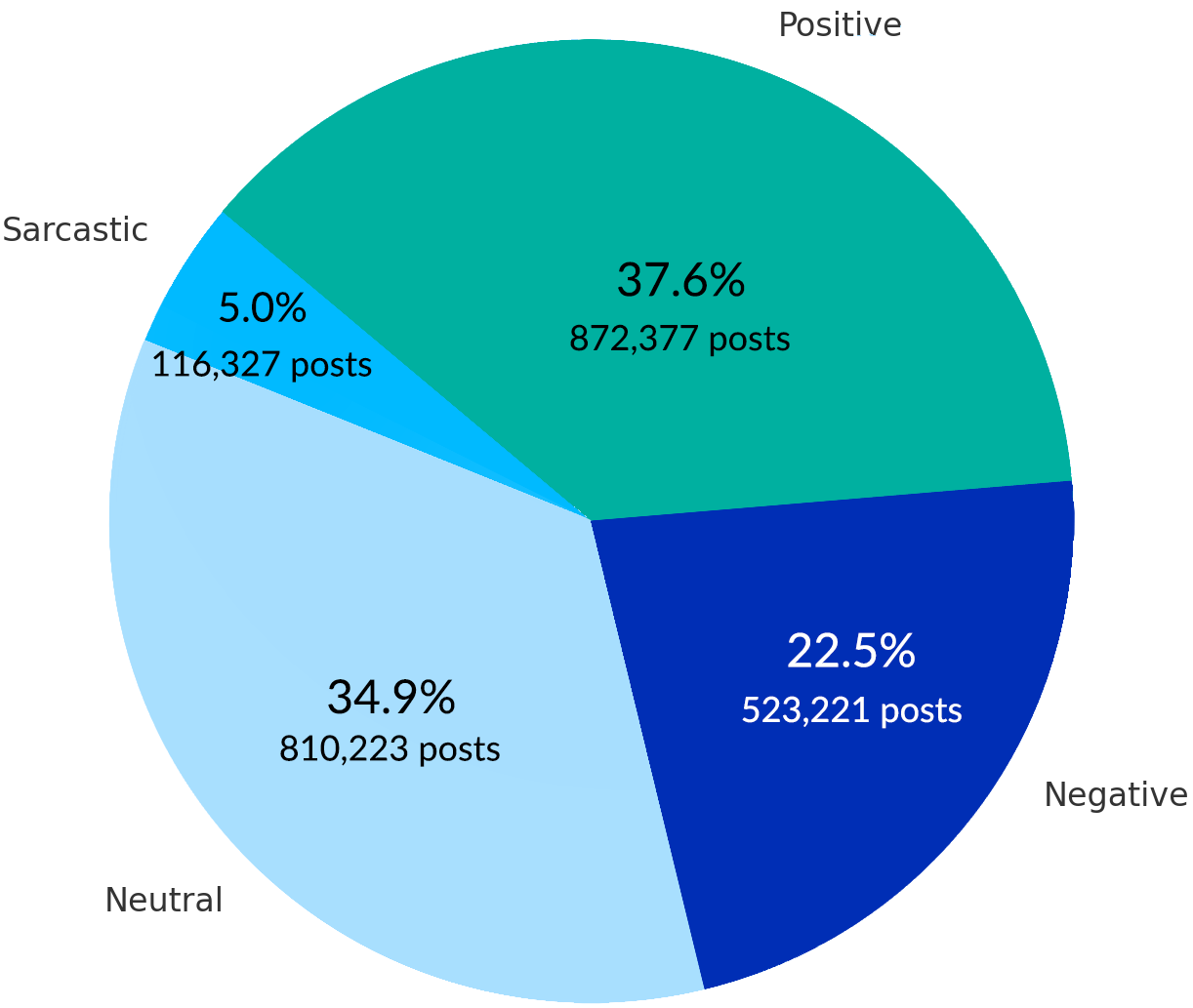}
        \label{piechart:b}}

\caption{Pie charts show the distribution of four sentiment categories (positive, neutral, negative, and sarcastic) among posts. Chart (a) represents the sentiment scale across all posts, while chart (b) focuses on distinct (unique) posts with no duplicates.}

\label{piechart}
\end{figure}

In the total sentiment distribution, the sentiment of 80\% of posts is neutral or positive, as seen in Figure~\ref{piechart:a}. However, if we discard the duplicated posts, the fraction of negative and sarcastic posts increases, and the neutral and positive posts drop to about 10\% of the total posts, indicating a larger scale of neutral and positive posts exist (Figure~\ref{piechart:b}).

\subsection{Comparing Trends of COVID-19 and Swine Fever}

\begin{figure}[H]


\centering

\includegraphics[width=\textwidth]{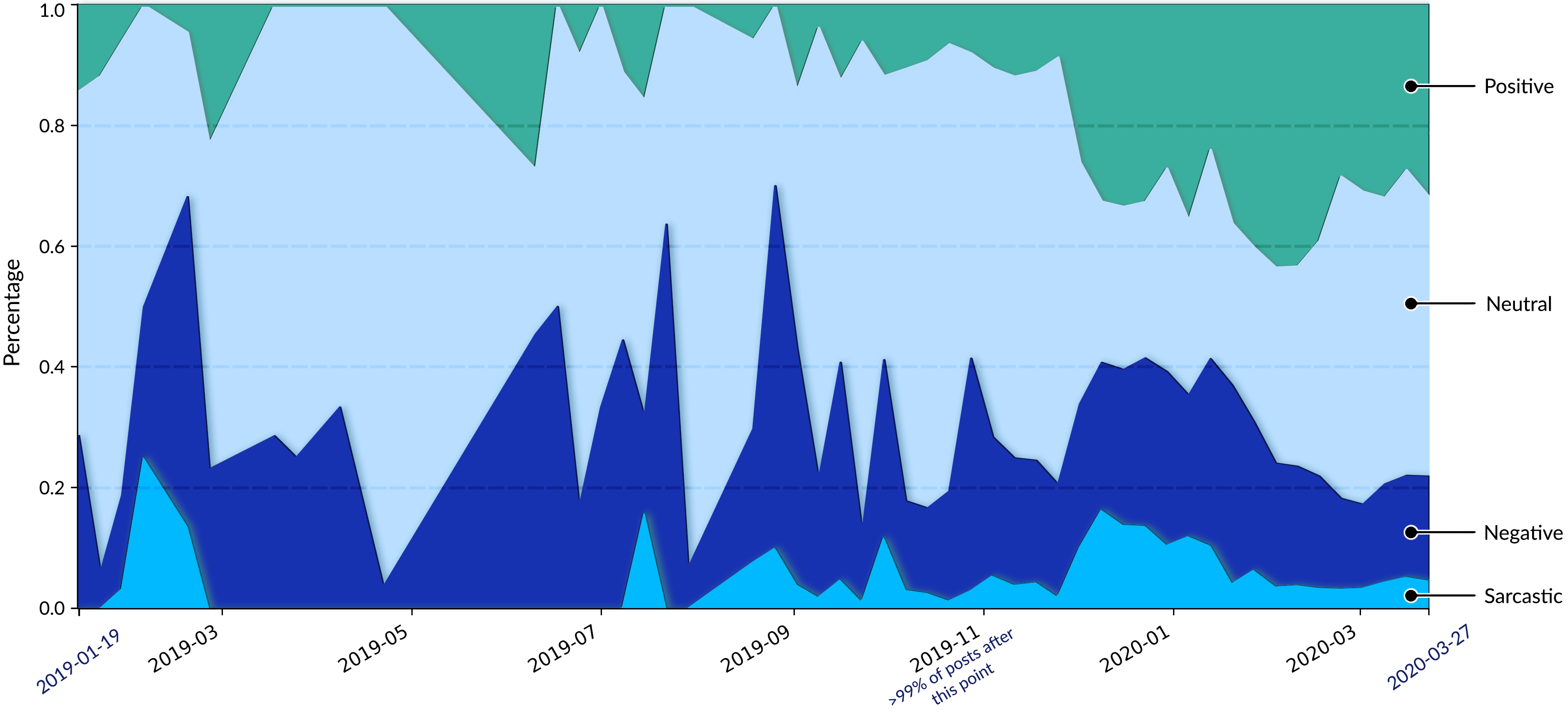}

\caption{Timeline stacked area chart showing the percentage distribution of four sentiments (positive, neutral, negative, sarcastic) in posts over time. The y-axis represents the current fraction of each sentiment at each time. Each segment's thickness corresponds to the proportion of that sentiment in the dataset. For example, a wider section indicates a higher percentage of that sentiment during the corresponding period on the x-axis. The x-axis represents days \textbf{from January 2019 to March 2020}, with key events or transitions, such as the note at '2019-11,' which marks where over 99\% of posts occurred after this point. This chart allows an easy visual comparison of the evolution of each sentiment over time.}

\label{stacked_chart}


\end{figure}

Figure~\ref{stacked_chart} shows the sentiment changes for distinct posts over time to reveal further the trends in the evolution of sentiment dynamics. There are two time periods during which there are significant spikes in negative post frequency, one in September 2019 and another in December 2019. 

The spike starting in December corresponds to the widely recognized outbreak of COVID-19. Analogically, the spike in negativity around September 2019 is attributable to the outbreak of African Swine Fever in South Korea. Given China's substantial number of South Korean restaurants and the interconnected food industry between the two countries, this outbreak triggered widespread public concern and panic on Weibo.

While these two negative spikes were due to disease outbreaks, they exhibit distinct patterns in terms of positive sentiment posts. In the case of September 2019, there is no noticeable increase in positive posts alongside the surge in negative posts. The overall tone on Weibo during this period remains predominantly negative and often sarcastic. In contrast, in December 2019, a substantial rise in positive sentiment posts was observed following the outbreak of COVID-19 in China. 

Several factors may contribute to these differing patterns. COVID-19 had a more significant impact on daily life in China than African Swine Fever had and fostered greater social cohesion and solidarity than less severe events due to this reason. This cohesion and solidarity led to a more significant presence of positive sentiment posts around the onset of the pandemic. From our analysis, two primary types of posts drove the rise in positive sentiment posts: (1) government efforts to emphasize the effectiveness of disease control measures, including on Weibo, and (2) spontaneous expressions of encouragement and mutual aid among Weibo users.

In contrast, during the September 2019 African Swine Fever outbreak, Weibo made no strong attempt to manage public speech about this disease, and this outbreak's impact on the daily lives of ordinary citizens in China was minimal. As a result, public discussion was full of rumors, such as concerns about the contamination of supermarket pork. The lack of authoritative communication and reassurance exacerbated public panic, intensifying the negative sentiment on Weibo compared to the outbreak of COVID-19.

\subsection{Synchronicity of Sentiment Trends}

Another notable observation is the synchronicity of dynamics across different sentiments. For instance, sarcastic sentiments and negative sentiments often exhibit synchronized patterns. A rise in sarcastic posts typically accompanies an increase in negative posts. This correlation is understandable, as sarcastic posts frequently convey negative opinions or criticisms.

Interestingly, this kind of correlation also occurs between positive and sarcastic sentiments, particularly during abrupt surges in positive sentiment posts triggered by events. In Figure~\ref{stacked_chart}, we can observe an apparent increase in positive sentiment posts starting in December 2019. 

Sarcastic sentiment posts deviate from their usual alignment with negative sentiments and instead rise alongside positive sentiments. A closer examination of these posts reveals that many sarcastic posts are directed at the surge in positive feelings, often satirizing the shift in societal tone. Some sarcastic posts reflect disputes among Weibo users, while others criticize the government. 

This phenomenon highlights that real-world events shape societal sentiments and influence one another. A surge in positive sentiments is often met with opposing voices, creating polarization within the online community. Such interactions underscore the complexity of sentiment dynamics in internet societies, where differing perspectives amplify societal divides. 

\subsection{Entropy Analysis of Sentiment Types}

\begin{table}[h!]
\captionsetup{justification=centering}
\caption{Sentiment Entropy and Distance}
\centering
\begin{tabular}{|c|c|c|c|c|c|}
\hline
\textbf{Type} & \textbf{Entropy} & \textbf{Rank} & \textbf{Distance} & \textbf{Pair Entropy} & \textbf{Pair Distance} \\
\hline
\cellcolor{orange!30}Neutral & \cellcolor{orange!30}$9.2740$ & \cellcolor{orange!30}Lowest & \cellcolor{orange!30}$0.0559$ & \cellcolor{orange!30} & \cellcolor{yellow!20}\\
\cline{1-4}
\cellcolor{orange!30}Positive & \cellcolor{orange!30}$9.3299$ & \cellcolor{orange!30}Low & \cellcolor{orange!30}$0.0785$ & \multirow{-2}{*}{\cellcolor{orange!30}$9.30195$} & \cellcolor{yellow!20}\\
\cline{1-5}
\cellcolor{yellow!30}Sarcastic & \cellcolor{yellow!30}$9.4084$ & \cellcolor{yellow!30}High & \cellcolor{yellow!30}$0.0483$ & \cellcolor{yellow!30} & \cellcolor{yellow!20}\\
\cline{1-4}
\cellcolor{yellow!30}Negative & \cellcolor{yellow!30}$9.4567$ & \cellcolor{yellow!30}Highest & \cellcolor{yellow!30}- & \multirow{-2}{*}{\cellcolor{yellow!30}$9.43255$} & \multirow{-4}{*}{\cellcolor{yellow!20}$0.1306$}\\
\cline{1-6}
\end{tabular}
\label{tab:entropy}
\end{table}

We compute the average entropy of posts for each sentiment type. Entropy measures the amount of information content per character, allowing us to compare the complexity of different sentiment categories. For this analysis, we sampled 15,000 posts from each sentiment type and calculated their average entropy in bits. Additionally, we computed the distance between each sentiment type by determining the difference between their entropy scores. The results are presented in Table~\ref{tab:entropy}, which details the average entropy for each type and the distances between them. We also paired the neutral and positive sentiments, as they are similar, and calculated their average entropy. Similarly, we paired sarcastic and negative sentiments and provided their average entropy as well. Lastly, we present the distance between these two pairs.

This novel method of using entropy allows us to uncover differences between various types of sentiments. Positive and neutral sentiments tend to have similar vocabularies, often derived from government posts, with few unique authors contributing. This results in a low entropy of words in these posts. In contrast, negative and sarcastic posts are typically authored by individuals who have directly experienced hardship due to recent events. Their expressions of suffering are personal and unique, leading to a higher entropy due to the limited overlap of vocabulary with other posts.

\section{Discussion}

Our study covers two novel contributions to sentiment analysis. We have conducted the first part of our analysis on a large-scale classifying sentiment of about three million Chinese language posts from the social network Weibo, using a local LLM model, Llama 3 8B. To guide the model, we wrote a few-shot learning prompt with examples of Chinese posts and their expected classifications selected by a native speaker. We established the model's accuracy using a validated set of 199 Chinese posts with ground-truth labels assigned by o1-mini and verified by a native speaker. We then cleaned the dataset and extracted 2,322,148 original posts, which we categorized using the model. Including the original posts and their duplicates, we classified 4,049,407 posts. 

Secondly, we analyzed the classified posts for sentiment trends on Weibo during the early stages of the COVID-19 pandemic from November 2019 to March 2020. Our analysis revealed more positivity in the online discussions of the COVID-19 outbreak than in the case of African Swine Fever outbreaking just a couple of months before. We also found a correlation between sarcastic post frequency and the frequency of spikes in positive or negative posts. The management of COVID-19 discourse by the government, as well as the polarization of opinions about the pandemic on Weibo, were factors that contributed to these trends. 

In summary, our methodology is effective for large-scale sentiment analysis of Chinese posts. An essential limitation of some conventional NLP tools is that they may not capture language-specific rhetorical devices, grammatical features, or slang that would indicate sentiment subtly, especially for non-English languages. Large language models can overcome this limitation as they can understand context-specific features. This understanding allows the model to distinguish sarcastic posts from negative or positive posts. 

Our work has applications as broad as other sentiment analysis methods. For governments and social media companies, our research shows the impact and resulting trends of public crises on social media.

One significant limitation of the methodology is that the F1-score for sarcastic post classification using Llama 3 (8B model) is low, approximately 50\%. Despite this, the model was still useful for identifying trends in sarcastic post patterns. Notably, the frequency of sarcastic posts within the COVID-19 dataset is relatively low. To enable larger-scale sarcasm detection, an improvement in accuracy is necessary, whether using Llama 3 or other large language models (LLMs). The language complexity of sarcastic posts, particularly those from users who faced hardships during the pandemic, likely contributes to this accuracy issue. Additionally, sarcastic postings make up a smaller proportion of the dataset compared to the other three types of posts.

Future work in this field is essential as the impact of sarcasm and other rhetorical devices, such as irony, on the accuracy of sentiment analysis is largely unexplored, especially in non-English datasets and social networks. One direction is to align large language models for more accurate detection of rhetorical devices in sentiment analysis use cases. As sentiment detection can involve subjective interpretations, future researchers may focus on the reasoning done by LLMs to make such classifications.

\authorcontributions{Conceptualization, B.K.S, C.J.H., and M.S.M.; methodology, C.J.H., and M.S.M.; validation, C.J.H., and M.S.M.; formal analysis, C.J.H., and M.S.M.; investigation, B.K.S, C.J.H., and M.S.M.; data curation, C.J.H.; writing—original draft preparation, B.K.S, C.J.H., and M.S.M.; writing review and editing, B.K.S, C.J.H. and M.S.M.; visualization, C.J.H., and M.S.M.; supervision, B.K.S.; project administration, B.K.S. and M.S.M.; funding acquisition, B.K.S. All authors have read and agreed to the published version of the manuscript.}

\funding{This research was partially funded by the National Science Foundation (NSF). Grant No. BSE-2214216}

\acknowledgments{The authors would like to thank RPI's Future of Computing Institute for providing computing resources that made this research possible.}

\reftitle{References}

\bibliography{bibliography}

\PublishersNote{}

\end{document}